\documentclass[notoc]{JHEP3}
\usepackage{oldgerm}

\font\fa=bbm11
\newcommand{\N}{\mbox{\fa N}}   

\newcommand{\Ibb}[1]{ {\rm I\ifmmode\mkern -3.6mu\else\kern -.2em\fi#1}}
\newcommand{\ibb}[1]{\leavevmode\hbox{\kern.3em\vrule
     height 1.2ex depth -.3ex width .2pt\kern-.3em\rm#1}}
\newcommand{\Cl}{{\ibb C}}           
\newcommand{\Rl}{{\Ibb R}}           

\newcommand{\Om}{\Omega}

\newcommand{\Hil}{\mathcal{H}}
\newcommand{\VV}{\mathcal{V}}

\newcommand{\Q}{\mathcal{Q}}
\newcommand{\F}{\mathcal{F}}

\newcommand{\DD}{\mathcal{D}} 
 
\newcommand{\W}{\mathcal{W}}    

\newcommand{\Ss}{\mathscr{S}}   
\newcommand{\OO}{O}   
\newcommand{\PGpo}{\mathcal{P}_+^\uparrow}   
\newcommand{\LGpo}{\mathcal{L}_+^\uparrow}
\newcommand{\LGo}{\mathcal{L}^\uparrow}
\newcommand{\LGa}{\mathcal{L}^\downarrow}

\newcommand{\PGo}{\mathcal{P}^\uparrow}
\newcommand{\LG}{\mathcal{L}}

\newcommand{\LGm}{\hat{\mathcal L}}
\newcommand{\PGm}{\hat{\mathcal P}}
 
\newcommand{\PG}{\mathcal{P}}

\newcommand{\DT}{\mathscr{D}}

\def\bp{{\mbox{\boldmath{$p$}}}}
\def\bq{{\mbox{\boldmath{$q$}}}}

\def\sbp{{\mbox{\scriptsize\boldmath{$p$}}}}

\newcommand{\supp}{\mathrm{supp}\,}

\newcommand{\vte}{\vartheta}
\newcommand{\la}{\lambda}
\newcommand{\La}{\Lambda}

\newcommand{\kae}{\kappa_{\rm e}}
\newcommand{\kam}{\kappa_{\rm m}}

\newcommand{\fti}{\widetilde{f}}

\newcommand{\fbar}{\overline{f}}

\newcommand{\zd}{z^{\dagger}}

\newcommand{\lto}{\longrightarrow}

\newcommand{\iin}{_\mathrm{in}}

\newcommand{\oout}{_\mathrm{out}}
\newcommand{\tp}[1]{^{\otimes #1}}    

\newcommand{\dom}{\mathrm{dom}}

\newcommand{\A}{\mathcal{A}}
\newcommand{\B}{\mathcal{B}}

\usepackage{amsmath}
\usepackage{amsfonts,amssymb}
\usepackage{mathrsfs}

\newtheorem{theorem}{Theorem}[section]
\newtheorem{proposition}[theorem]{Proposition}
\newtheorem{lemma}[theorem]{Lemma}

\newtheorem{definition}[theorem]{Definition}

\title{Wedge-Local Quantum Fields and Noncommutative Minkowski Space}
\author{Harald Grosse\\ Faculty of Physics, University of Vienna,\\ Boltzmanngasse 5, A-1090 Vienna, Austria\\ E-mail: \email{harald.grosse@univie.ac.at}}
\author{Gandalf Lechner\\ International Erwin Schr\"odinger Institute for Mathematical Physics,\\ Boltzmanngasse 9, A-1090 Vienna, Austria,\\ E-mail: \email{gandalf.lechner@esi.ac.at}}

\abstract{
Within the setting of a recently proposed model of quantum fields on noncommutative Minkowski space, the consequences of the consistent application of the proper, untwisted Poincar\'e group as the symmetry group are investigated. The emergent model contains an infinite family of fields which are labelled by different noncommutativity parameters, and related to each other by Lorentz transformations. The relative localization properties of these fields are investigated, and it is shown that to each field one can assign a wedge-shaped localization region in Minkowski space. This assignment is consistent with the principles of covariance and locality, i.e. fields localized in spacelike separated wedges commute. 

Regarding the model as a non-local, but wedge-local, quantum field theory on ordinary (commutative) Minkowski spacetime, it is possible to determine two-particle S-matrix elements, which turn out to be non-trivial. Some partial negative results concerning the existence of observables with sharper localization properties are also obtained.
}

\begin{document}

\section{Introduction}

In relativistic quantum field theories, Einstein causality is implemented by requiring that the observables of spacelike separated observers are represented by commuting operators. This principle of locality is usually assumed to hold for arbitrarily small spacelike distances. However, all current approaches to quantum physics trying to incorporate effects of quantum gravity, like string theory \cite{Eliezer:1989cr}, quantum field theory on noncommutative spacetimes \cite{Szabo:2001kg,Douglas:2001ba} or loop quantum gravity \cite{Thiemann:2002nj}, show some kind of non-local behaviour. In fact, in these theories locality is usually a meaningful concept only in some large scale limit.

The probably simplest examples of such theories are quantum field theories on a deformed, noncommutative Minkowski space, on which the coordinates $\hat{x}_\mu$ satisfy a commutation relation of the form
\begin{align}\label{xmn1}
 [\hat{x}_\mu, \hat{x}_\nu]=i\,Q_{\mu\nu}\,.
\end{align}
Here the noncommutativity parameter $Q$ is some real, antisymmetric $(d\times d)$-matrix. Spacetime models of this form can be motivated by considering the restrictions on event measurements suggested by classical gravity and the uncertainty principle \cite{Doplicher:1994tu}, or, in certain cases, as low-energy limits of string theory \cite{Seiberg:1999vs}.

In this paper, we study a specific model on noncommutative Minkowski space from a new point of view, and investigate its locality properties, which turn out to be quite different from what is usually expected.

As our starting point we take in Section \ref{sec:nc} the scalar massive free field $\phi(Q_1,x)$ on noncommutative Minkowski space, with a fixed noncommutativity parameter $Q_1$. Formulating this field as an operator on Fock space, we then generate a whole family of non-local quantum fields $\phi(Q,x)$ from it by acting on $\phi(Q_1,x)$ with the usual second quantized representation $U$ of the proper Poincar\'e group associated to the free scalar field of mass $m>0$.

The emerging model has similarities with other systems studied in the literature: For fixed $Q$, the $n$-point functions of the field $\phi(Q,x)$ coincide with the $n$-point functions recently proposed by Fiore and Wess \cite{Fiore:2007vg} and Chaichian et.~al. \cite{Chaichian:2004za}. However, we do not use a twisted version of the Poincar\'e group as these authors do. Rather, the use of a representation of the untwisted symmetry forces us to consider a large class of fields $\phi(Q,x)$, labelled by an orbit of noncommutativity parameters, and related to each other by Lorentz transformations.

Our model has also connections to the free field model studied by Doplicher, Fredenhagen and Roberts in \cite{Doplicher:1994tu}, since also there, a whole spectrum of noncommutativity parameters is considered. However, due to a different action of the Poincar\'e group, there exist also essential differences between the two models, which are spelled out in detail in Section \ref{sec:nc}.

In Section \ref{sec:wedges}, we prove our main result stating that the fields $\phi(Q,x)$, albeit non-local, are far from being completely delocalized with respect to each other. These relative locality properties are proven with the help of a novel construction, associating to each $Q$ an infinitely extended, wedge-shaped spacetime region $W\subset\Rl^d$. We define a bijection $W\mapsto Q(W)$ between a set $\W_0$ of wedges and a set $\Q$ of noncommutativity parameters, and consider the corresponding fields $\phi_W(x):=\phi(Q(W),x)$. It is then shown that this association of spacetime regions to field operators is completely consistent with the principles of covariance and locality: 

Under the adjoint action of the representation $U$, they transform according to
\begin{align}
 U(y,\La)\phi_W(x)U(y,\La)^{-1} &= \phi_{\La W}(\La x+y)\,,\qquad (y,\La)\in\PGpo\,.
\end{align}
In $d=4$ dimensions, we make the well-known observation that each field $\phi_W$ alone transforms covariantly only under the subgroup SO$(1,1)\times{\rm SO}(2)$ of the Lorentz group. However, the whole family $\{\phi_W\,:\,W\in\W_0\}$ respects the full Lorentz symmetry.

Furthermore, for two wedges $W,\tilde{W}\in\W_0$ and two spacetime points $x,y$, we find (wedge-) local commutativity in the form
\begin{align}
 [\phi_W(x),\phi_{\tilde{W}}(y)]=0\qquad {\rm if}\;\,(W+x)\;\;{\rm is\,\;spacelike\,\;to}\;\;(\tilde{W}+y)\,.
\end{align}
Therefore $\phi_W(x)$ may be interpreted as a field configuration localized in the region $W+x$ instead of the point set $\{x\}$. Such a type of localization is similar to the string-local quantum fields studied by Mund, Schroer and Yngvason \cite{Mund:2005cv}, but different from the usual uniform nonlocality of quantum fields on noncommutative Minkowski space. 

This difference in localization is due to the fact that in comparison to usual quantum field theories on Minkowski space with fixed noncommutativity $Q$, we consider here a model encompassing a whole Lorentz orbit of noncommutativites, and study the relations of the corresponding subtheories with respect to each other.
\\\\
The fields $\phi_W(x)$ introduced here can also be understood as wedge-local quantum fields on genuine, ``commutative'' Minkowski space $\Rl^d$, $d\geq 2$. From this point of view, our analysis fits into the ongoing research aiming at constructions of quantum field theory models with the help of wedge-localized operators \cite{Schroer:1997cq,Lechner:2003ts,GL07-1,Buchholz:2004qy,Longo:2004fu,Buchholz:2005xj}. In this context, such fields are mostly used as auxiliary quantities, which, due to their weakened locality properties, can be constructed more easily than point-local Wightman fields, also in the presence of non-trivial interactions \cite{GL07-1}.

It is shown in Section \ref{sec:qft} that from this perspective, the model defined by the fields $\phi_W$ has a number of similarities to completely integrable quantum field theories in two dimensions. In particular, the algebra of the creation and annihilation parts of the free field on noncommutative Minkowski space is quite similar \cite{Kulish:2006jq} to the Zamolodchikov-Faddeev algebra \cite{Zamolodchikov:1978xm}, which underlies integrable models with factorizing S-matrices \cite{smirnov-book}. Making use of the model-independent results about wedge-local operators found by Borchers, Buchholz and Schroer \cite{Borchers:2000mz}, we calculate two-particle S-matrix elements and show that the model under consideration describes non-trivial interaction.

In any spacetime dimension $d\geq 2$, we therefore arrive at a theory of wedge-local, interacting quantum fields. This observation requires an investigation of the question if there exist also observables with sharper localization properties, like localization in bounded spacetime regions, in this setting. First results indicating that there probably is no strictly local quantum field theory corresponding to our wedge-local construction are presented in Section \ref{sec:qft}.

The article ends in Section \ref{sec:end} with a discussion of the results and an account of open questions.

\section{Noncommutative Minkowski spacetime and twisted CCR algebras}\label{sec:nc}

The simplest example of a noncommutative spacetime is the noncommutative counterpart of Minkowski space (of dimension $d\geq 2$), which is usually described by a ${}^*$-algebra of selfadjoint coordinate operators $\hat{x}_\mu$, $\mu=0,...,d-1$, satisfying
\begin{align}\label{xmunu}
 [\hat{x}_\mu,\hat{x}_\nu]=i\,Q_{\mu\nu}\,,
\end{align}
with a fixed, real, antisymmetric $(d\times d)$-matrix $Q$ \cite{Douglas:2001ba}, called the {\em noncommutativity parameter} or simply the {\em noncommutativity}.

Considering the most interesting four-dimensional case, this algebra can be realized in its Schr\"odinger representation \cite{Doplicher:1994tu}, where the $\hat{x}_\mu$ are operators on (a dense domain in) $L^2(\Rl^2,d^2s)$,
\begin{align*}
 (\hat{x}_0 \psi)(s_1,s_2) &= \kappa_{\rm e}\, s_1\cdot \psi(s_1,s_2)\,,
\qquad& 
(\hat{x}_1 \psi)(s_1,s_2) &= -i(\partial_{s_1}\psi)(s_1,s_2)\,,\\
(\hat{x}_2 \psi)(s_1,s_2) &= \kappa_{\rm m}\, s_2\cdot \psi(s_1,s_2)\,,
\qquad& (\hat{x}_3 \psi)(s_1,s_2) &= -i(\partial_{s_2}\psi)(s_1,s_2)\,.
\end{align*}
Here $\kae$ and $\kam$ are arbitrary real parameters measuring the strength of noncommutative effects, and the matrix $Q$ takes its standard form
\begin{align}\label{standardQ4d}
 Q = 
 \left(
\begin{array}{cccc}
 0 & \kae & 0 & 0\\
 -\kae & 0  & 0 & 0\\
 0        & 0  & 0 & \kam\\
 0 & 0    & -\kam & 0
\end{array}
\right)\,.
\end{align}
If $Q\neq 0$, the relations \eqref{xmunu} are not invariant under the natural action of the full Lorentz group on the vector $\hat{x}$, and therefore Lorentz covariance is broken from the outset if $Q$ is taken to be a fixed matrix. Recently, proposals were made how to formulate a ``twisted'' action of the Lorentz group in this setting, leading to a different concept of covariance \cite{Fiore:2007vg,Chaichian:2004za,Akofor:2007hk}.

However, such a deformation of the symmetry group is not necessary if one allows for a richer operator form of the commutator $-i[\hat{x}_\mu,\hat{x}_\nu]$, encompassing a whole spectrum of numerical matrices $Q$. This approach was taken by Doplicher, Fredenhagen and Roberts and shown to lead to models with better covariance properties \cite{Doplicher:1994tu}.

In the present article, we work in a somewhat similar framework, and consider a family of quantum fields $\phi(Q,x)$ which depend explicitly on the noncommutativity parameter $Q$, and are related to each other by Lorentz transformations.
\\\\
To describe the construction of such fields and their relations to similar models studied in the literature, let us introduce some notation. We consider the free scalar quantum field $\phi_0$ of mass $m>0$ on ordinary (``commutative'') $d$-dimensional Minkowski spacetime, $d\geq 2$. The energy of a particle with momentum $\bp\in\Rl^{d-1}$ is denoted $\omega_\sbp:=(\bp^2+m^2)^{1/2}$, and the upper mass shell by $H_m^+:=\{(\omega_\sbp,\bp)\,:\,\bp\in\Rl^{d-1}\}$. Generally, we shall use the letters $p,q$ for on-shell momenta, and the boldface letters $\bp,\bq\in\Rl^{d-1}$ for their respective spatial components.

Defined as an operator-valued distribution, $\phi_0$ acts on its domain in the Bosonic Fock space $\Hil=\bigoplus_{n=0}^\infty\Hil_n$ over the single particle space $\Hil_1:=L^2(H^+_m,d\mu)$, where $d\mu:=d^{d-1}\bp/\omega_\sbp$ is the Lorentz invariant measure on $H^+_m$.

On $\Hil$, we have the usual (anti-) unitary second quantized representation $U$ of the Poincar\'e group with spin zero and mass $m$. The proper orthochronous transformations $(y,\La)\in\PGpo$, and the total reflections $j_\mu$ mapping $x_\mu$ to $-x_\mu$ and leaving the other coordinates unchanged, are represented as ($\Psi\in\Hil$, $p_1,...,p_n\in H^+_m$)
\begin{subequations}\label{def:U}
\begin{align}
 (U(y,\La)\Psi)_n(p_1,...,p_n)     &= e^{i\sum_{l=1}^n p_l\cdot y}\cdot\Psi_n(\La^{-1}p_1,...,\La^{-1}p_n)\,,\\
 (U(0,j_0)\Psi)_n(p_1,...,p_n)     &= \overline{\Psi_n(-j_0p_1,...,-j_0p_n)}\,,\label{defUj}\\
 (U(0,j_k)\Psi)_n(p_1,...,p_n)     &= \Psi_n(j_kp_1,...,j_kp_n)\,,\qquad k=1,...,d-1\,.
\end{align}
\end{subequations}
In particular, the total spacetime reflection $j:=j_0\cdots j_{d-1}:x\mapsto-x$ acts by complex conjugation on each $n$-particle space.

The generators of the translations $U(y_\mu,1)$ will be denoted by $P_\mu$, and the corresponding vacuum vector by $\Omega\in\Hil$.

The free field is defined with the help of the standard representation of the CCR algebra on $\Hil$, i.e. we have the creation/annihilation operators $a^*(p),a(p)$, $p\in H^+_m$, which satisfy
\begin{align}
  a(p)a(q)   &= a(q)a(p)\,,\\
  a(p)a^*(q) &= a^*(q)a(p)+\omega_\sbp\,\delta(\bp-\bq)\, {\rm id}_\Hil\,.
\end{align}
These operators give the field
\begin{align}\label{def:phi0}
\phi_0(x) := \int d\mu(p)\,\bigg(e^{ip\cdot x}a^*(p) + e^{-ip\cdot x} a(p)\bigg)\,,
\end{align}
which after smearing in $x$ becomes an unbounded operator on $\Hil$ containing the dense subspace $\DD\subset\Hil$ of finite particle number in its domain.
\\\\
As is well known, $\phi_0$ has a counterpart on noncommutative Minkowski space, which can be realized on the tensor product $\VV\otimes\Hil$ of the representation space $\VV$ of the coordinate operators $\hat{x}_\mu$ and Fock space $\Hil$ as \cite{Doplicher:1994tu}
\begin{align}\label{def:phixhat}
 \phi_\otimes(Q,x) = \int d\mu(p)\left(e^{ip\cdot x} a^*_\otimes(Q,p) + e^{-ip\cdot x} a_\otimes(Q,p) \right)\,,
\end{align}
with the creation/annihilation operators
\begin{align}\label{ahat}
 a_\otimes(Q,p) := e^{-ip\cdot \hat{x}}\otimes a(p)\,,\qquad  a^*_\otimes(Q,p) := e^{ip\cdot \hat{x}}\otimes a^*(p)\,,
\end{align}
taking values in the operators on $\VV\otimes\Hil$. Here $\hat{x}_\mu$ satisfies \eqref{xmunu} for some arbitrary $Q\in\Rl_{d\times d}^-$, the space of real, antisymmetric $(d\times d)$-matrices. We indicate the dependence on $Q$ explicitly in our notation, since $Q$ will be allowed to vary later on.

As a consequence of \eqref{xmunu}, the $a^\#_\otimes(Q,p)$ satisfy the commutation relations
\begin{subequations}\label{algahat}
\begin{align}
 a_\otimes(Q,p)a_\otimes(Q,p') 	&= e^{-ipQp'} a_\otimes(Q,p')a_\otimes(Q,p)\,,\qquad pQp':=p_\mu Q^{\mu\nu}p'_\nu\,,\\
 a^*_\otimes(Q,p)a^*_\otimes(Q,p') 	&= e^{-ipQp'} a^*_\otimes(Q,p')a^*_\otimes(Q,p)\,,\\
 a_\otimes(Q,p)a^*_\otimes(Q,p')&= e^{+ipQp'} a^*_\otimes(Q,p')a_\otimes(Q,p) + \omega_\sbp\delta(\bp-\bp')\,{\rm id}_{\VV\otimes\Hil}\,.
\end{align}
\end{subequations}
It has been realized by a number of authors (see, for example, \cite{Akofor:2007hk,Grosse-ZZ}) that this algebra can also be represented on $\Hil$ instead of $\VV\otimes\Hil$ by using the following distributions, containing the energy-momentum operators $P_\mu$,
\begin{equation}\label{def:aQp}
 a(Q,p) := e^{\frac{i}{2}pQP}a(p)\,,\qquad a^*(Q,p) := e^{-\frac{i}{2}pQP}a^*(p)\,.
\end{equation}
Using
\begin{align}
 e^{\frac{i}{2}pQP}\,a(p') = e^{-\frac{i}{2}pQp'}\cdot a(p')\,e^{\frac{i}{2}pQP}
\end{align}
and the antisymmetry of $Q$, it is easy to show that $a^*(Q,p)=a(Q,p)^*$ and that the $a^\#(Q,p)$ also satisfy the relations \eqref{algahat}.

Let us denote the corresponding field operators by
\begin{align}\label{def:phiQ}
 \phi(Q,x) &:= \int d\mu(p)\left(e^{ip\cdot x}\, a^*(Q,p) + e^{-ip\cdot x}\, a(Q,p) \right)\,.
\end{align}
In the context of these or similar fields, some authors propose to work with a ``twisted'' Poincar\'e algebra \cite{Fiore:2007vg,Chaichian:2004za,Akofor:2007hk} to arrive at a covariant formulation despite $Q$ being constant. We take here a different point of view and and use the well-known representation $U$ \eqref{def:U} of the untwisted Poincar\'e group to implement the relativistic symmetry.

Since the adjoint action of $U$ on the field $\phi(Q,x)$ induces also a transformation of $Q$ (see Lemma \ref{lemma:a+U} below), it is necessary to consider a whole family of fields labelled by noncommutativity parameters. It is thus of interest to determine the commutation relations between the commutation/annihilation operators $a^\#(Q,p)$ and $a^\#(Q',p')$ for $Q\neq Q'$, generalizing \eqref{algahat}. For a somewhat related discussion, see \cite{Fiore:2007vg}.

By straightforward calculation, one finds the following exchange relations, valid for arbitrary on-shell momenta $p,p'\in H^+_m$ and  matrices $Q,Q'\in\Rl_{d\times d}^-$:
\begin{align}\label{aQp}
 a(Q,p)a(Q',p') &= e^{-\frac{i}{2}p(Q+Q')p'}\,a(Q',p')a(Q,p)\,,\nonumber\\
 a^*(Q,p)a^*(Q',p') &= e^{-\frac{i}{2}p(Q+Q')p'}\,a^*(Q',p')a^*(Q,p)\,,\\
 a(Q,p)a^*(Q',p') &= e^{\frac{i}{2}p(Q+Q')p'}\,a^*(Q',p')a(Q,p) + \omega_\sbp\,\delta(\bp-\bp')\,e^{\frac{i}{2}p(Q-Q')P}.\nonumber
\end{align}
Starting from this ``twisted'' CCR algebra, we consider the quantum fields $\phi(Q,x)$ \eqref{def:phiQ}, depending not only on the spacetime points $x\in\Rl^d$ (in the sense of distributions), but also on the matrices $Q$.

In the remainder of this section we analyze the transformation behaviour of these fields under Poincar\'e transformations. Afterwards, a comparison to other models \cite{Doplicher:1994tu,Fiore:2007vg,Chaichian:2004za} will be presented.
\\\\
For the trivial noncommutativity parameter $Q=0$, we see that $\phi(0,x)=\phi_0(x)$ coincides with the free field on commutative Minkowski space. Consequently, $\phi(0,x)$ enjoys the well-known covariance and locality properties of a Wightman field.

If $Q\neq 0$, however, the field $\phi(Q,x)$ is neither local nor does it transform covariantly under the full Lorentz group. To see its nonlocality explicitly, we compute the two-particle contribution of the field commutator $[\phi(Q,x),\phi(Q,y)]$ applied to the vacuum. Using the antisymmetry of $Q$, we find
\begin{align}
 2i\int d\mu(p)\int d\mu(q)\,e^{i(px+qy)}\,\sin\left(\frac{q_\mu Q^{\mu\nu}p_\nu}{2}\right)a^*(p)a^*(q)\Om\,.
\end{align}
This expression does not vanish for spacelike separated $x,y$ except for the case $Q=0$.
\\
\\
To study the transformation behaviour of the fields $\phi(Q,x)$ under Poincar\'e transformations, we consider the action of the Poincar\'e group on the algebra \eqref{aQp}. 

It is shown in the lemma below that the adjoint action of $U(0,\La)$ on the $a^\#(Q,p)$, where $\La$ is an element of the Lorentz group $\LG$, induces on $Q$ the transformation
\begin{align}\label{def:gamma}
Q\longmapsto\gamma_\La(Q) &:= \left\{\begin{array}{rcl}
                                  \La Q\La^T &;&\La\in\LGo\\
		                 -\La Q\La^T &;&\La\in \LGa
                         \end{array}
\right.\;\;,\qquad Q\in\Rl_{d\times d}^-\,.
\end{align}
Here, as usual, $\LGo$ and $\LGa$ denote the sets of orthochronous and anti-orthochronous Lorentz transformations, respectively, and corresponding notations are used for the associated subsets of the Poincar\'e group. Note that in view of the structure of the Lorentz group, $\gamma$ is an $\LG$-action, i.e. $\gamma_{\La\La'}=\gamma_\La\gamma_{\La'}$, $\La,\La'\in\LG$.
\\
\begin{lemma}\label{lemma:a+U}{\bf (Transformation properties of the twisted CCR algebra)}\\
The operator-valued distributions $a^\#(Q,p)$, $Q\in\Rl_{d\times d}^-$, $p\in H^+_m$, transform under the adjoint action of $U$ \eqref{def:U} according to, $y\in\Rl^d$, $\La\in\LG$,
\begin{align}
 U(y,\La)a^*(Q,p)U(y,\La)^{-1}		&=	e^{\pm i\La p\cdot y}\,a^*(\gamma_\La(Q),\pm\La p)\,,\label{acov1}\\
 U(y,\La)a(Q,p)U(y,\La)^{-1}		&=	e^{\mp i\La p\cdot y}\,a(\gamma_\La(Q),\pm\La p)\,,\label{acov2}
\end{align}
where the first sign is valid for $\La\in\LGo$ and the second sign holds for $\La\in\LGa$. Hence the fields $\phi(Q,x)$ \eqref{def:phiQ} satisfy
\begin{align}\label{phi-trans}
 U(y,\La)\phi(Q,x)U(y,\La)^{-1}	&= \phi(\gamma_\La(Q),\La x +y)\,,\qquad (y,\La)\in\PG\,.
\end{align}
\end{lemma}
\Proof
We begin by looking at orthochronous Poincar\'e transformations $(y,\La)\in\PGo$ and find
\begin{align*}
U(y,\La)a^*(Q,p) U(y,\La)^{-1}
&= e^{i\La p\cdot y}U(y,\La)e^{-\frac{i}{2}p_\mu Q^{\mu\nu} P_\nu}U(y,\La)^{-1}a^*(\La p)\\ 
&= e^{i\La p\cdot y}\,e^{-\frac{i}{2}(\La p)_\mu (\La Q\La^T)^{\mu\nu} P_\nu}a^*(\La p)\\
&=
e^{i\La p\cdot y}\,a^*(\La Q\La^T,\La p)
=
e^{i\La p\cdot y}\,a^*(\gamma_\La(Q),\La p)\,.
\end{align*} 
In complete analogy, one shows \eqref{acov2} for $(y,\La)\in\PGo$. For transformations involving time reflection  $j_0$, we take into account that $U(0,j_0)$ \eqref{defUj} is a conjugate linear operator, which leads to a change of sign in the exponent of $\exp(\pm\frac{i}{2}pQP)$. Since $U(0,j_0)a^*(p)U(0,j_0)^{-1}=a^*(-j_0p)$, time reflection acts on $a^*(Q,p)$ according to
\begin{align*}
 U(0,j_0)a^*(Q,p)U(0,j_0)^{-1} &= e^{+\frac{i}{2}(-j_0p)_\mu (j_0Qj_0^T)^{\mu\nu}P_\nu}a^*(-j_0p)
= a^*(\gamma_{j_0}(Q),-j_0p)
\,,
\end{align*} 
as claimed in \eqref{acov1}. The argument for \eqref{acov2} is completely analogous, and the transformation behaviour \eqref{phi-trans} of $\phi(Q,x)$ follows directly from (\ref{acov1}, \ref{acov2}).
{\hfill  $\square$ \\[2mm]}

In the following proposition, we show that despite the violation of the usual covariance and locality properties of quantum field theory, the fields $\phi(Q,x)$ satisfy the remaining Wightman axioms \cite{streater-wightman} for a scalar field. Furthermore, the Reeh-Schlieder property (cyclicity of the vacuum for the field algebra, see \cite{streater-wightman}) holds.

We consider the fields smeared with Schwartz testfunctions $f\in\Ss(\Rl^d)$, i.e. the distributions
\begin{align}\label{phidist}
 f\longmapsto\phi(Q,f)&=\int d^dx\,f(x)\phi(Q,x) = a^*(Q,f^+)+a(Q,f^-)\,,\\
a^\#(Q,f^\pm)&:=\int d\mu(p)f^\pm(p)a^\#(Q,p)\,,\\
		 f^\pm(p) &:= \int d^dx\,f(x)\,e^{\pm ip\cdot x}\,,\qquad p=(\omega_\sbp,\bp)\in H^+_m\,.
\end{align}

\begin{proposition}\label{prop:phi}{\bf(Wightman properties of the field operators $\phi(Q,f)$)}\\
Let $Q\in\Rl_{d\times d}^-$ and $f\in\Ss(\Rl^d)$.
\begin{enumerate}
 \item The dense subspace $\DD$ of vectors of finite particle number is contained in the domain of any $\phi(Q,f)$ and is stable under the action of these operators.
\item For $\Psi\in\DD$, 
\begin{align}
 f\longmapsto \phi(Q,f)\Psi
\end{align}
is a vector-valued tempered distribution.
\item For $\Psi\in\DD$ one has
\begin{align}
 \phi(Q,f)^*\Psi &= \phi(Q,\fbar)\Psi\,,
\end{align}
and for real $f$, $\phi(Q,f)$ is essentially selfadjoint on $\DD$.
\item The Reeh-Schlieder property holds: For any non-empty open subset $\OO\subset\Rl^d$,
\begin{align}
 \DT_Q(\OO)	&:= {\rm span}\,\{\phi(Q,f_1)\cdots\phi(Q,f_n)\Om\;:\;n\in\N_0,\,f_1,...,f_n\in\Ss(\OO)\}
\end{align}
is dense in $\Hil$.
\end{enumerate}
\end{proposition}
\Proof
Since the proofs of these statements are very similar to the corresponding arguments for the well-known free field $\phi_0$, we can be brief about them.

a) The fact that $\DD\subset\dom\,\phi(Q,f)$ and $\phi(Q,f)\DD\subset\DD$ follows directly from the definition of this operator \eqref{phidist}.

b) Since the operators $e^{\pm\frac{i}{2}pQP}$ give only rise to multiplication by phases, the smeared creation/annihilation operators satisfy the familiar bounds, $n\in\N_0$,
\begin{align}
 \|a^\#(Q,f^\pm)\Psi_n\| \leq \sqrt{n+1}\cdot\|\Psi_n\|\left(\int d\mu(p)\,|f^\pm(p)|^2\right)^{1/2}\,,\qquad\Psi_n\in\Hil_n\,.
\end{align}
In the topology of $\Ss(\Rl^d)$, the right hand side depends continuously on $f$. Hence the temperateness of $f\longmapsto \phi(Q,f)\Psi$, $\Psi\in\DD$, follows.

c) The first statement is a straightforward consequence of the fact that $Q$ is real. The essential selfadjointness for real $f$ can be proven along the same lines as \cite[Prop. 5.2.3]{BraRob2} by showing that every $\Psi\in\DD$ is an entire analytic vector for $\phi(Q,f)$.

d) Making use of the spectral properties of the representation $U(y,1)$ of the translations, one can apply the standard Reeh-Schlieder argument \cite{streater-wightman} to show that $\DT_Q(\OO)$ is dense in $\Hil$ if and only if $\DT_Q(\Rl^d)\subset\Hil$ is dense. Choosing $f_1,...,f_n\in\Ss(\Rl^d)$ such that the supports of the Fourier transforms $\fti_k$ do not intersect the lower mass shell, one notes that $\DT_Q(\Rl^d)$ contains the vectors
\begin{align*}
\phi(Q,f_1)\cdots\phi(Q,f_n)\Om=
 a^*(Q,f^+_1)\cdots a^*(Q,f^+_n)\Om &= \sqrt{n!}\,P_n(S_n(f^+_1\otimes...\otimes f^+_n))\,,
\end{align*}
where $P_n$ denotes the orthogonal projection from $\Hil_1\tp{n}$ onto its totally symmetric subspace $\Hil_n$, and $S_n\in\B(\Hil_1\tp{n})$ is the operator multiplying with
\begin{align}
 S_n(p_1,...,p_n) &= \prod_{1\leq l<k\leq n}e^{-\frac{i}{2}p_lQp_k}\,.
\end{align}
Varying the testfunctions $f_k\in\Ss(\Rl^d)$ within the above limitations gives rise to dense sets of one particle wavefunctions $f_k^+$ in $\Hil_1$. Since $S_n$ is a unitary operator on $\Hil_1\tp{n}$, the vectors $S_n(f^+_1\otimes...\otimes f^+_n)$ which can be obtained in this way form a total set in $\Hil_1\tp{n}$. Hence the image of this set under the projection $P_n$ is total in $\Hil_n$, which implies the density of $\DT_Q(\Rl^d)\subset\Hil$ in view of the Fock structure of $\Hil$.
{\hfill  $\square$ \\[2mm]}

Before we analyze the localization properties of the fields $\phi(Q,x)$ in the following sections, we point out some relations to other models studied in the literature.

In the recent preprint \cite{Fiore:2007vg}, Fiore and Wess consider a framework for quantum field theories on noncommutative Minkowski space (with fixed noncommutativity $Q$), in which coordinate differences are represented by commuting operators. This leads them to consider $n$-point functions of the form
\begin{align}\label{npt}
{\mathscr W}^Q_n(x_1,...,x_n) := \langle\Om\,,\phi(x_1)\star_Q\cdots\star_Q\phi(x_n)\,\Om\rangle\,,
\end{align}
where $\phi$ denotes a Wightman field on commutative Minkowski space and $\star_Q$ is a Moyal-like product. For $f\in\Ss(\Rl^{n\cdot d}), g \in\Ss(\Rl^{m\cdot d})$ (and analogously, for tempered distributions), $f\star_Q g$ is defined as
\begin{align}
 (f\star_Q g)(x_1,...,x_n,y_1,...,y_m) = \prod_{l=1}^n\prod_{k=1}^m\exp\left(-\frac{i}{2}\frac{\partial}{\partial x_l^\mu}Q^{\mu\nu}\frac{\partial}{\partial y_k^\nu}\right) f(x_1,...,x_k)g(y_1,...,y_l).\nonumber
\end{align}
In a more ad hoc manner, the same $n$-point functions have also been proposed in \cite{Chaichian:2004qk}.	

Taking in \eqref{npt} for $\phi$ the free field $\phi_0$, and considering a fixed $Q$, the exchange relations of the $a^\#(Q,p)$ imply
\begin{align}
\phi(Q,x_1)\cdots\phi(Q,x_n)\,\Om
&=
\prod_{1\leq l<k\leq n}\exp\left(-\frac{i}{2}\frac{\partial}{\partial x_l^\mu}Q^{\mu\nu}\frac{\partial}{\partial x_k^\mu}\right)\,\phi_0(x_1)\cdots\phi_0(x_n)\,\Om
\,.\nonumber
\end{align}
A proof of this equation can be carried out using induction in the field number $n$. In particular, the vacuum expectation values of the field products $\phi(Q,x_1)\cdots\phi(Q,x_n)$ coincide with the distributions ${\mathscr W}^Q_n$ \eqref{npt}.

This observation clarifies the meaning of our model as a theory containing the fields studied by Fiore and Wess \cite{Fiore:2007vg} and Chaichian et. al. \cite{Chaichian:2004za} as subtheories, for different values of the noncommutativity $Q$. Put differently, what we analyze in this article are not the individual subtheories given by fixed $Q$, but rather the {\em relative} properties of these fields with respect to each other.
\\
\\
We would also like to compare our fields $\phi(Q,x)$ to the model formulated in terms of the tensor product field operators $\phi_\otimes(Q,x)$ \eqref{def:phixhat} \cite{Doplicher:1994tu} in more detail.

As mentioned before, the creation and annihilation operators $a_\otimes^\#(Q,p)$ \eqref{ahat} and $a^\#(Q,p)$ \eqref{def:aQp} satisfy the same algebraic relations if $Q$ is fixed \eqref{algahat}. More precisely, the ${}^*$-algebra $\F^Q_\otimes$ generated by the smeared fields $\phi_\otimes(Q,f)$, $f\in\Ss(\Rl^d)$, is isomorphic to the ${}^*$-algebra $\F^Q$ generated by the fields $\phi(Q,f)$. One can easily calculate that this isomorphism is implemented by the following unitary operator.

Given $Q\in\Rl_{d\times d}^-$ and $\xi\in\VV$, $\|\xi\|_\VV=1$, we define $V_{Q,\xi}^{(n)}:\Hil_n\to\VV\otimes\Hil_n\cong L^2((H^+_m)^{\times n}\to\VV,d\mu^{\times n})$,
\begin{align}
 (V_{Q,\xi}^{(n)}\Psi_n)(p_1,...,p_n) &:= \Psi_n(p_1,...,p_n)\cdot e^{i(p_1+...+p_n)\hat{x}}\xi\,,\qquad\Psi_n\in\Hil_n\,.
\end{align}
The direct sum of these operators, $V_{Q,\xi}:=\bigoplus_{n=0}^\infty V_{Q,\xi}^{(n)}$, is a unitary mapping $\Hil$ to $\VV\otimes\Hil$, and 
\begin{align}
 V_{Q,\xi} \,\F^Q V_{Q,\xi}^* &= \F^Q_\otimes\,,\qquad
 V_{Q,\xi} \,a^\#(Q,p)V_{Q,\xi}^* = a_\otimes^\#(Q,p)\,,
\end{align}
where the second equation holds in the sense of distributions.

In particular, it follows from $V_{Q,\xi}\Om=\xi\otimes\Om$ that the following $n$-point functions coincide, $x_1,...,x_n\in\Rl^d$,
\begin{align}\label{wm1}
\langle(\xi\otimes\Om),\,\phi_\otimes(Q,x_1)\cdots\phi_\otimes(Q,x_n)\,(\xi\otimes\Om)\rangle
&=
\langle\Om,\,\phi(Q,x_1)\cdots\phi(Q,x_n)\,\Om\rangle\,.
\end{align}
So in the context of the fields $\phi_\otimes(Q,x)$, we are working with a vector state of the form $\langle\xi\otimes\Om,\,.\;\xi\otimes\Om\rangle_{\VV\otimes\Hil}$. The choice of $\xi\in\VV$ is irrelevant as long as $\|\xi\|=1$, since $\langle\xi,\,.\,\xi\rangle_\VV$ is evaluated only on the identity operator in $\B(\VV)$ due to momentum conservation in the second tensor factor.

The use of such a state differs drastically from the approach taken in \cite{Doplicher:1994tu}, where the concepts of noncommutative geometry were applied to identify pure states on the spacetime algebra as the noncommutative analogues of points in the undeformed spacetime manifold. This analogy suggests to consider different states $\xi$ on each factor $\phi_\otimes(Q,x_k)$, in contrast to \eqref{wm1}.

Besides this different choice of state, there are also differences in the algebraic structure between the fields considered here and the fields considered in \cite{Doplicher:1994tu}, if different values of $Q$ are taken into account. In \cite{Doplicher:1994tu}, the action of a Lorentz transformation $\La$ on the coordinate operators is simply given by $\hat{x}^\mu\mapsto\La^\mu_{\;\,\nu}\hat{x}^\nu$. The transformed $\hat{x}_\mu$ therefore have the commutator matrix $\La Q\La^T$, which coincides with $\gamma_\La(Q)$ for orthochronous $\La$.

However, the commutation relations of the transformed creation/annihilation operators are different from those found in \eqref{acov1}. For example,
\begin{align}
 \big( e^{ip\cdot\La\hat{x}}\otimes a^*(p)\big)
 \big( e^{ip'\cdot \hat{x}}\otimes a^*(p')\big)
&=
e^{-\frac{i}{2}p(\La Q)p'}\cdot
\big( e^{ip'\cdot \hat{x}}\otimes a^*(p')\big) 
\big( e^{ip\cdot\La\hat{x}}\otimes a^*(p)\big)\,,
\end{align}
instead of \eqref{aQp}, which gives the exchange factor $e^{-\frac{i}{2}p(Q+\La Q\La^T)p'}$. Put differently, the isomorphism $\F^Q_\otimes\cong\F^Q$ mentioned before does not carry over to the algebras $\F_\otimes:=\bigvee_Q\F^Q_\otimes$ and $\F:=\bigvee_Q\F^Q$ generated by $\F^Q_\otimes$ and $\F^Q$ for different noncommutativity parameters $Q$.

\section{Wedges and wedge-local quantum fields}\label{sec:wedges}

We now set out to analyze the localization properties of the fields $\phi(Q,x)$ \eqref{def:phiQ}. As has been shown before, these are non-local fields in the sense that they violate the usual point-like localization of Wightman fields. However, we will argue that they are not completely delocalized, either: It turns out that $\phi(Q,x)$ is localized in an infinitely extended, wedge-shaped region of Minkowski space.

The wedge $\phi(Q,x)$ is localized in depends on both, the spacetime point $x$ and the matrix $Q$. To establish the wedge-locality of these fields, the essential step is to construct a map $W\longmapsto Q(W)$ from a set $\W_0$ of wedges (defined below) to a set $\Q\subset\Rl_{d\times d}^-$ of antisymmetric matrices such that $\phi_W(x):=\phi(Q(W),x)$ is localized in the shifted wedge $W+x$. This construction is carried out in the first Subsection \ref{ss:sym}.

Afterwards, in Subsection \ref{ss:w}, the concept of covariant, wedge local quantum fields will be introduced, and the properties of the operators $\phi_W(x)=\phi(Q(W),x)$ will be explained.

To avoid confusion, we emphasize that the wedges considered here have no relation to the so-called ``lightwedge'' sometimes mentioned in the literature \cite{AlvarezGaume:2003mb}.

\subsection{Symmetries of wedges and noncommutativity parameters}\label{ss:sym}

To define a correspondence between noncommutativity parameters $Q$ and localization regions, we first recall some well-known definitions and facts about {\em wedges} in $d$-dimensional Minkowski space.

As our reference wedge region, we take
\begin{equation}\label{def:w1}
 W_1 := \{x\in\Rl^d\,:\,x_1>|x_0|\}\,,
\end{equation}
often called the {\em right wedge} in the literature. With respect to the coordinate reflections $j_\mu:x_\mu\mapsto-x_\mu$, this region has the symmetry properties
\begin{align}\label{w1-sym}
 j_0W_1=W_1\,,\qquad j_1W_1=-W_1=W_1'\,,\qquad j_kW_1=W_1\,,\quad k>1\,,
\end{align}
where $W_1'$ denotes the causal complement of $W_1$.

The {\em set $\W$ of all wedges} in $\Rl^d$ is defined as the set of all Poincar\'e transforms of $W_1$, i.e. $\W:=\PG W_1$. We will mostly work with a subset $\W_0\subset\W$, consisting only of the {\em Lorentz} transforms of $W_1$,
\begin{equation}
 \W_0 := \LG W_1 \subset\W\,.
\end{equation}
$\W_0$ contains precisely all those wedges which have the origin in their edges, see also \cite{Buchholz:1998pv,Kuckert:2005na}. Note that in view of the symmetries \eqref{w1-sym}, and since $j$ commutes with $\LG$, there holds
\begin{align}\label{w-w}
 W'=jW=-W\,,\qquad W\in\W_0\,.
\end{align}
Moreover, it follows from \eqref{w1-sym} that for spacetime dimension $d>2$,
\begin{eqnarray}
\W_0=\LGpo W_1\,,\qquad (d>2)\,.
\end{eqnarray}
Hence $(\W_0,\iota)$ is a homogeneous space for the proper orthochronous Lorentz group $\LGpo$ when equipped with the natural action $\iota$,
\begin{align}\label{def:iota}
 \iota_\La(W):=\La W\,.
\end{align}
As is well known, the corresponding stabilizer group $\LGpo(W_1,\iota)\subset\LGpo$ of $W_1$ is $\LGpo(W_1,\iota)={\rm SO}(1,1)\times{\rm SO}(d-2)$, where the ${\rm SO}(1,1)$ factor contains the Lorentz boosts in $x_1$-direction and ${\rm SO}(d-2)$ describes rotations in the edge of $W_1$, i.e. in the subspace $\{x\in\Rl^d\,:\,x_0=x_1=0\}$.\footnote{Whereas the invariance of $W_1$ under spatial rotations around the $x_1$-axis is obvious from \eqref{def:w1}, the SO$(1,1)$-invariance follows from the fact that a boost in $x_1$-direction has the lightlike vectors $(\pm 1,1,0,...,0)$ as eigenvectors, with positive eigenvalues.
}

In the following, we will consider the subgroup $\LGm\subset\LG$ which is generated by the proper orthochronous part $\LGpo$ of $\LG$ and the total spacetime reflection $j:x\mapsto -x$. (In even spacetime dimensions, $\LGm$ coincides with the proper Lorentz group $\LG_+$, while in odd dimensions, $\LGm=\LGpo\cup\LG^\downarrow_-$.) The corresponding subgroup of the Poincar\'e group will be denoted $\PGm:=\LGm\rtimes\Rl^d$.

Considering $\W_0$ as a homogeneous space for the larger group $\LGm\supset\LGpo$, the stabilizer $\LGm(W_1,\iota)$ of $W_1$ arises from $\LGpo(W_1,\iota)$ by adjoining the reflection $j_0j_2$ if $d>2$ is even and by adjoining $j_0$ if $d>2$ is odd.

In the two-dimensional case, all transformations in $\LGpo={\rm SO}(1,1)$ leave $W_1$ invariant, and hence
\begin{align}
 \W_0 = \{W_1,-W_1\}\,,\qquad (d=2)\,.
\end{align}
This two-element set forms an $\LGm$-homogeneous space, too, and we have the stabilizer groups $\LGpo(W_1,\iota)=\LGm(W_1,\iota)=\LGpo$ for $d=2$.

The various stabilizers are collected in Table 1.

\TABLE[ht]{
\begin{tabular}{|l||l|l|}
\hline
 dimension $d$ & $\LGpo(W_1,\iota)$&$\LGm(W_1,\iota)$\\
 \hline\hline
 $d>2$ even & ${\rm SO}(1,1)\times{\rm SO}(d-2)$ & $\LGpo(W_1,\iota)\cup j_0j_2\LGpo(W_1,\iota)$\\
 $d>2$ odd  & ${\rm SO}(1,1)\times{\rm SO}(d-2)$ & $\LGpo(W_1,\iota)\cup j_0\LGpo(W_1,\iota)$\\
 $d=2$      & $\LGpo$                            & $\LGpo$\\
 \hline
\end{tabular}
\caption{Stabilizer groups of $W_1$ \eqref{def:w1} with respect to the action $\iota$ \eqref{def:iota}.} 
}

To set up the desired correspondence $W\longmapsto Q(W)$ between wedges and antisymmetric matrices, we construct in the following an $\LGm$-homogeneous space $(\Q,\gamma)$ which is isomorphic to $(\W_0,\iota)$. Since we want to understand the covariance properties \eqref{phi-trans} of the fields $\phi(Q,x)$, we want to endow $\Q$ with the action $\gamma$ \eqref{def:gamma}, restricted to $\LGm\subset\LG$, i.e.
\begin{align}\label{def:gamma2}
Q\longmapsto\gamma_\La(Q) &:= \left\{\begin{array}{rcl}
                                  \La Q\La^T &;&\La\in\LGpo\\
		                 -\La Q\La^T &;&\La\in j\LGpo
                         \end{array}
\right.\;\;,\qquad Q\in\Rl_{d\times d}^-\,.
\end{align}
We do not take the full Lorentz group as our symmetry group here because, as is shown below, a homomorphism $(\W_0,\iota)\cong(\Q,\gamma)$ does not exist in the most important four-dimensional case if these spaces are treated as homogeneous spaces for all of $\LG$.

As a homogeneous space, $\Q$ must consist of a single $\LGm$-orbit under $\gamma$, i.e. there exists $Q_1\in\Rl_{d\times d}^-$ such that
\begin{align}
 \Q := \{\gamma_\La(Q_1)\,:\,\La\in\LGm\,\}\,.
\end{align}
The matrix $Q_1$ will be chosen in such a way that the map (also denoted $Q$)
\begin{eqnarray}\label{mapQ}
  Q:\W_0\lto\Q\,,\qquad Q(\La W_1):=\gamma_\La(Q_1)\,,\qquad\La\in\LGm\,,
\end{eqnarray}
is well-defined. If this is the case, \eqref{mapQ} implies that $Q$ is a homomorphism of $\LGm$-homogeneous spaces, i.e. it intertwines the actions $\iota$ and $\gamma$, $Q\circ\iota_\La=\gamma_\La\circ Q$, $\La\in\LGm$.

The possible choices of $Q_1$ and the resulting properties of $\Q$ are discussed in the following lemma.
\begin{lemma}\label{lemma:QW}
\begin{enumerate}
 \item The mapping $Q$ \eqref{mapQ} is a homomorphism of the $\LGm$-homogeneous spaces $(\W_0,\iota)$ and $(\Q,\gamma)$ if and only if $Q_1$ has the form (depending on the spacetime dimension $d$)
\begin{subequations}\label{Q1}
\begin{align}
 Q_1
&=
\left(
\begin{array}{ccccc}
 0 & \kae & 0 & \cdots & 0\\
 -\kae & 0  & 0 & \cdots & 0\\
 0      & 0  & 0 & \cdots & 0\\
 \vdots & \vdots &\vdots&\ddots&\vdots\\
0 & 0&0&\cdots&0
\end{array}
\right)\,,& (d&\neq 4)\,,\label{Q1dn4}\\
 Q_1
&=
\left(
\begin{array}{cccc}
 0 & \kae & 0 & 0\\
 -\kae & 0  & 0 & 0\\
 0        & 0  & 0 & \kam\\
 0 & 0    & -\kam & 0
\end{array}
\right)\,,&(d&=4)\,,\label{Q1d4}
\end{align}
\end{subequations}
with arbitrary $\kae, \kam\in\Rl$. If $Q_1\neq 0$, $Q$ is actually an isomorphism, i.e. invertible.
\item If $Q_1$ has the form \eqref{Q1}, its $\LGm$-orbit $\Q$ is
\begin{align}
 \Q &= \{Q_1,-Q_1\}\,,& (d&=2)\,,\label{setQ2}\\
 \Q &= \{\La Q_1\La^T\,:\, \La\in\LGpo\}\,,&(d&>2)\,,\label{setQg}
\end{align}
and for any $W\in\W_0$, there holds
\begin{equation}\label{Qprime}
 Q(W')	= -Q(W) = \gamma_j(Q(W))\;.
\end{equation}
\end{enumerate}
\end{lemma}
\Proof
 a) We begin with the proof of the ``only if'' statement. Clearly, the map \eqref{mapQ} is only well-defined if the stabilizer group $\LGm(W_1,\iota)$ is contained in the stabilizer $\LGm(Q_1,\gamma)$ of $Q_1$ with respect to the action $\gamma$. In particular, $Q_1$ must satisfy (cf. Table 1)
 \begin{align}
  \La Q_1\La^T=Q_1\,,\qquad \La\in {\rm SO}(1,1)\times{\rm SO}(d-2)\,.
 \end{align}
Hence it must be of the block form 
\begin{align}\label{q1-d}
Q_1
=
\left(
\begin{array}{ccccc}
 0 & \kae & 0 & \cdots & 0\\
 -\kae & 0  & 0 & \cdots & 0\\
 0        & 0 \\
 \vdots & \vdots &&\tilde{Q}_1\\
0 & 0
\end{array}
\right)\,,
\end{align}
where $\kae\in\Rl$ is an arbitrary parameter and $\tilde{Q}_1\in \Rl_{(d-2)\times(d-2)}^-$ is rotationally invariant, i.e. 
\begin{align}
 \tilde{Q}_1=R\tilde{Q}_1 R^T = R\tilde{Q}_1R^{-1}\,,\qquad R\in{\rm SO}(d-2)\,.
\end{align}
If $d=2$, the block $\tilde{Q}_1$ is absent, and in case $d=3$, it must vanish due to the antisymmetry of $Q$. In $d=4$, $\tilde{Q}_1=\left(\begin{array}{cc}0&\kam\\-\kam&0\end{array}\right)$, $\kam\in\Rl$, is the most general form of an antisymmetric $(2\times 2)$-matrix. Finally, if $d>4$ and $\tilde{Q}_1\neq 0$, the linear span of $\tilde{Q}_1$ defines a non-trivial ${\rm SO}(d-2)$-invariant subspace of $\Rl_{(d-2)\times(d-2)}^-$. Since the adjoint action of ${\rm SO}(n)$ on $\Rl_{n\times n}^-$ is irreducible for $n>2$, we conclude $\tilde{Q}_1=0$ for $d>4$, i.e. we have shown that \eqref{Q1} is necessary for $Q$ \eqref{mapQ} to be well-defined.

Now we consider the ``if'' part, i.e. assume that $Q_1$ is of the form \eqref{Q1}. If $Q_1=0$, the map $Q(W)=0$ is clearly well-defined, but trivial and not invertible. 

So assume $Q_1\neq 0$. At first, one computes that the stabilizer subgroup $\LGpo(Q_1,\gamma)$ coincides with $\LGpo(W_1,\iota)$ in this case: $\LGpo(W_1,\iota)\subset\LGpo(Q_1,\gamma)$ is guaranteed by the choice \eqref{Q1} of $Q_1$, and equality of the two stabilizers follows by explicit inspection of the action of boosts and rotations on \eqref{Q1}, see also \cite{Doplicher:1994tu}.

Since $\gamma_j(Q_1)=-Q_1$ \eqref{def:gamma2} and $\LGm$ is generated by $\LGpo$ and $j$, it follows that $\LGm(Q_1,\gamma)$ is generated by $\LGpo(Q_1,\gamma)$ and
\begin{align}\label{-g}
 \{\La\in\LGpo\,:\,\La Q_1\La^T=-Q_1\}\,.
\end{align}
Using the special form of $Q_1$, it can be easily shown that $j_0j_2$ respectively $j_0$ belongs to \eqref{-g} if the spacetime dimension $d>2$ is even and odd, respectively, whereas in $d=2$, \eqref{-g} is empty.

It follows from these considerations that in any dimension $d$, we have
\begin{align}
 \LGm(Q_1,\gamma) = \LGm(W_1,\iota)\,,
\end{align}
which is equivalent to the map $Q$ being invertible.

b) Since $\gamma_j(Q_1)=-Q_1$, we have in general 
\begin{align}
\Q=\{\La Q_1\La^T\,:\,\La\in\LGpo\} \cup \{-\La Q_1\La^T\,:\,\La\in\LGpo\}\,.
\end{align}
In $d=2$ dimensions, $\LGm(Q_1,\gamma)=\LGpo$, and therefore $\Q$ reduces to $\Q=\{Q_1,-Q_1\}$ in this case \eqref{setQ2}. For the case that the spacetime dimension is larger than 2, one computes $(j_1j_2)Q_1(j_1j_2)^T=-Q_1$, which implies \eqref{setQg} since $j_1j_2\in\LGpo$.

For the proof of the statement \eqref{Qprime}, recall that $jW=W'$ for any $W\in\W_0$ \eqref{w-w}. Hence
\begin{align}
 Q(W') = Q(jW) = \gamma_j(Q(W)) = -Q(W)\,,
\end{align}
as claimed in \eqref{Qprime}.
{\hfill  $\square$ \\[2mm]}

This lemma shows that in $d=4$ and $d=2$, the symmetries of the standard noncommutativity parameter \eqref{standardQ4d} correspond precisely to the symmetries of the right wedge $W_1$ \eqref{def:w1}. If $d\neq 4$, or $d=4$ and $\kam=0$, the map $Q$ is also an isomorphism between $\W_0$ and $\Q$ if these are considered as homogeneous spaces for the {\em full} Lorentz group $\LG$, i.e. if we consider the unrestricted $\LG$-action \eqref{def:gamma} on $\Q$. However, in $d=4$ dimensions with $\kam\neq 0$, this is not the case, since for example $j_2W_1=W_1$, but $\gamma_{j_2}(Q_1)\neq Q_1$.

As a comparison to the analysis in \cite{Doplicher:1994tu}, let us point out that in $d=4$, the orbit $\Q$ can be explicitly characterized by two Lorentz invariant quantities as
\begin{align}
 \Q &= \{Q\in\Rl_{4\times 4}^-\,:\,Q_{\mu\nu}Q^{\mu\nu}=2(\kam^2-\kae^2),\;Q_{\mu\nu}(*Q)^{\mu\nu}=4\,\kae\kam\}\,,&(d&=4)\,,
\end{align}
i.e. for $|\kae|=|\kam|=1$, we have in the terminology of \cite{Doplicher:1994tu} $\Q=\Sigma_\pm$, where the sign $\pm$ is the sign of $\kae\cdot\kam$.

If $d\notin\{2,4\}$, the matrix $Q_1$ associated to $W_1$ has not the form expected from considerations on non-commutative Minkowski space, but is always of rank $2$, i.e. there must exist a large subspace of commuting coordinates $\hat{x}_2,...,\hat{x}_{d-1}$ for the correspondence between $\W_0$ and $\Q$ to hold also in these (physically less interesting) cases.

\subsection{A class of wedge-local model theories}\label{ss:w}

For the discussion of the fields $\phi(Q,x)$ we introduce the concept of a {\em $\PGm$-covariant, wedge-local quantum field}. A closely related notion of {\em string-localized} fields can be found in \cite{Mund:2005cv}.

For a model-independent definition, we consider some separable Hilbert space $\Hil$ with an (anti-) unitary positive energy representation $U$ of $\PGm$ with unique vacuum acting on it, and a dense, $U$-stable domain $\DD\subset\Hil$. 
Let $\phi$ denote a family $\{\phi_W\,:\,W\in\W_0\}$ of fields that satisfy the domain and continuity assumptions of Wightman's axioms \cite{streater-wightman}. The covariance and locality axioms are adopted to the wedge-local setting as follows.
\begin{definition}\label{defwlqf}
In the setting indicated above, $\phi=\{\phi_W\,:\,W\in\W_0\}$ is defined to be a wedge-local quantum field transforming covariantly under $U$ if the following two conditions are satisfied:
\begin{itemize}
 \item {\bf Covariance:} For any $W\in\W_0$ and $f\in\Ss(\Rl^d)$ there holds
\begin{align}
 U(y,\La)\phi_W(f)U(y,\La)^{-1} &= \phi_{\La W}(f\circ(y,\La)^{-1})\,,\qquad (y,\La)\in\PGpo\,,\label{phicov}\\
 U(0,j)\phi_W(f)U(0,j)^{-1} &= \phi_{jW}(\fbar\circ j)\,.\label{jonphi}
\end{align}
 \item {\bf Wedge-Locality:} Let $W,\tilde{W}\in\W_0$ and $f,g\in\Ss(\Rl^2)$. If
\begin{align}\label{Wsupp}
 \overline{W+\supp f} \subset (\tilde{W}+\supp g)'\,,
\end{align}
then 
\begin{align}\label{phi-comm}
 [\phi_W(f),\phi_{\tilde{W}}(g)]\Psi&=0\,,\qquad\Psi\in\DD\,.
\end{align}
\end{itemize}
\end{definition}

The condition \eqref{Wsupp} states that the localization region $W+\supp f$ of the field $\phi_W$, smeared with the testfunction $f$, should be spacelike separated from the localization region $\tilde{W}+\supp g$ of $\phi_{\tilde{W}}(g)$, for these operators to commute. Due to geometrical properties of the family of wedges, this condition can also be stated in a somewhat simpler form. To do so, we first recall the well-known fact that for an inclusion $W\subset\tilde{W}$ of wedges $W,\tilde{W}\in\W$, one necessarily has $W=\tilde{W}+a$ for some $a\in\Rl^d$ \cite{Thomas:1997ni,Buchholz:1998pv}. Moreover, if $\tilde{W}\in\W_0$, the Lorentz equivalence of this wedge to $W_1$ can be used to show that $a$ must be contained in the closure of $\tilde{W}$.

Consider two wedges $W,\tilde{W}\in\W_0$ and two compact sets $\OO,\tilde{\OO}\subset\Rl^d$ such that $\overline{W+\OO}\subset(\tilde{W}+\tilde{\OO})'$. For arbitrary points $x\in\OO$, $\tilde{x}\in\tilde{\OO}$, we then have the chain of inclusions
\begin{align}
\overline{W}+x \subset \overline{W +\OO} \subset (\tilde{W}+\tilde{\OO})'\subset \tilde{W}'+\tilde{x}\,.
\end{align}
From the remark made above, and from the fact that wedges are open, it now follows that $W=\tilde{W}'=-\tilde{W}$ and $x-\tilde{x}\in W$. Hence there exists a translation $a\in\Rl^d$ such that $\OO+a\subset W$, $\tilde{\OO}+a\subset -W=W'$. In other words, the condition $\overline{W+\OO}\subset(\tilde{W}+\tilde{\OO})'$ is Poincar\'e equivalent to $W=\tilde{W}'$, $\OO\subset W$, $\tilde{\OO}\subset W'$.

This observation leads us to the following lemma.
\begin{lemma}\label{lemma:geo}
 Let $\phi=\{\phi_W\,:\,W\in\W_0\}$ be a collection of fields satisfying the above regularity assumptions and the covariance properties stated in Def. \ref{defwlqf}. Then $\phi$ is wedge local if and only if
\begin{align}\label{commre}
 [\phi_{W_1}(f),\phi_{-W_1}(g)]\Psi=0\,,\qquad\Psi\in\DD\,,
\end{align}
for all $f,g\in C_0^\infty(\Rl^d)$ with $\supp f\subset W_1$ and $\supp g\subset-W_1$.
\end{lemma}
\Proof
The previous geometrical consideration implies that if \eqref{commre} holds for arbitrary compact $\supp f\subset W_1$, $\supp g \subset -W_1$, then \eqref{phi-comm} holds for all $W,\tilde{W}\in\W_0$ and $f,g \in C_0^\infty(\Rl^d)$ such that $\overline{W+\supp f}\subset (\tilde{W}+\supp g)'$.

Since each $\phi_W$ is assumed to be a temperate distribution, the extension to the larger class of Schwartz testfunctions then follows from the continuity of $f\mapsto\langle\Psi',\phi_W(f)\Psi\rangle$, $\Psi'\in\DD$, and the density of $\DD$ in $\Hil$.
{\hfill  $\square$ \\[2mm]}
Having clarified the geometry of causal configurations of wedges, we now switch again to the concrete setting of the previously introduced fields $\phi(Q,x)$. With the help of the homomorphism $Q:\W_0\to\Q$ \eqref{mapQ} studied in Lemma \ref{lemma:QW}, we define 
\begin{align}\label{defphi-2}
 \phi_W(x) := \phi(Q(W),x)=\int d\mu(p)\,\Big(a^*(Q(W),p)\,e^{ip\cdot x} + a(Q(W),p)\,e^{-ip\cdot x}\Big)\,.
\end{align}
It is the aim of the present section to show that the collection $\{\phi_W\}_{W\in\W_0}$ is a wedge-local quantum field in the sense of Definition \ref{defwlqf}.
\\\\
As before, these fields have to be understood as distributions, i.e. we consider the ``smeared'' fields (cf. \eqref{phidist})
\begin{align}\label{phif}
 \phi_W(f)	&= a^*(Q(W),f^+) + a(Q(W),f^-)\,.
\end{align}
As has been shown in Proposition \ref{prop:phi}, the fields $\phi_W$ comply with the assumptions on domain and continuity properties preceding Definition \ref{defwlqf}.

\begin{proposition}\label{prop:phiW}
If $\kae\geq 0$ in \eqref{Q1}, the collection of fields $\phi=\{\phi_W\,:\,W\in\W_0\}$ \eqref{phif} is a wedge-local quantum field on the Fock space $\Hil$, transforming covariantly under the representation $U$ \eqref{def:U} of $\PGm$.
\end{proposition}
\noindent{\em Remark:} If $\kae<0$, the homomorphism \eqref{mapQ} between $\Q$ and $\W_0$ has to be changed by a sign for the statement of the proposition to hold.
\\\\
\Proof
{\em Covariance.} Using the transformation properties \eqref{phi-trans} and the intertwining property $Q(\La W)=\gamma_\La(Q(W))$ \eqref{mapQ}, we immediately see, $(y,\La)\in\PGpo$,
\begin{align}
 U(y,\La)\phi_W(f)U(y,\La)^{-1} &= \phi(\gamma_\La(Q(W)),f\circ(y,\La)^{-1})= \phi_{\La W}(f\circ(y,\La)^{-1})\,.
\end{align}
Since $U(0,j)$ is a conjugate linear operator, $f$ has to be replaced by its complex conjugate $\fbar$, for this equation to hold for transformations $(y,\La)\in j\PGpo$.

{\em Wedge-Locality}. In view of Lemma \ref{lemma:geo}, it is sufficient to consider the wedge $W=W_1$ and compactly supported test functions $f\in C_0^\infty(W_1)$, $g\in C_0^\infty(-W_1)$. Taking into account $Q(W_1')=-Q(W_1)=-Q_1$ \eqref{Qprime} and the commutation relations \eqref{aQp}, the field commutator \eqref{commre} simplifies to
\begin{align*}
 [\phi_{W_1}(f),\phi_{-W_1}(g)]\Psi &= \left([a(Q_1,f^-), a^*(-Q_1,g^+)]+[a^*(Q_1,f^+), a(-Q_1,g^-)]\right)\Psi \,.
\end{align*}
Evaluation of the $n$-particle contribution of this vector at $q_1,...,q_n\in H^+_m$ gives
\begin{align}\label{phicomm}
([\phi_{W_1}(f),\phi_{-W_1}&(g)]\Psi)_n(q_1,...,q_n)\\
&=
\int d\mu(p)\left(e^{ipQ_1q} f^-(p) g^+(p)
-
e^{-ipQ_1q} f^+(p) g^-(p)\right)\cdot\Psi_n(q_1,...,q_n)\,,\nonumber\\
{\rm with}\quad q&:= \sum_{k=1}^n q_k\,.
\end{align}
In the following, we are going to show that the above integral vanishes for any $q$. To this end, we introduce new coordinates:
\begin{align}
 m_\perp  &:= (m^2+p_\perp^2)^{1/2}\,,\qquad p_\perp:=(p_2,...,p_{d-1})\,,\qquad
\vte := {\rm Arsinh}\frac{p_1}{m_\perp}\,.
\end{align} 
In the coordinates $(\vte,p_\perp)$, the integration measure and on-shell momentum are
\begin{align}
d\mu(p)= \frac{d^{d-1}\bp}{\omega_\sbp} = d\vte\,d^{d-2}p_\perp\,,\qquad 
p = p(\vte) := 
\left(
\begin{array}{cc}
 m_\perp\cosh\vte\\ m_\perp\sinh\vte \\ p_\perp
\end{array}
\right)\,.
\end{align} 
The correspondingly transformed functions $f^\pm$ are denoted by
\begin{align}
 f^\pm_{p_\perp}(\vte) &:= f^\pm(m_\perp\sinh\vte,p_\perp) = (2\pi)^{d/2}\,\fti(\pm p(\vte))\,.
\end{align} 
Since $f$ has compact support, its Fourier transform $\fti$ is entire analytic, and as $\sinh$ is entire, too, so are the functions $\vte\longmapsto f^\pm_{p_\perp}(\vte)$, $p_\perp\in\Rl^{d-2}$. To estimate the values of these functions for complex $\vte$, note that ($x\in\Rl^d$)
\begin{align}
 {\rm Im}\,p(\vte+i\la)\cdot x = m_\perp\sin\la 
\left(
\begin{array}{cc}
 \sinh\vte\\ \cosh\vte\\ 0
\end{array}
\right)
\cdot
\left(
\begin{array}{cc}
 x_0\\x_1\\x_\perp
\end{array}
\right)
 \,,\qquad x_\perp:=(x_2,...,x_{d-1})\,.
\end{align}
Assuming $x\in\supp f\subset W_1$, the two-vector $(x_0,x_1)$ lies in the two-dimensional right wedge, and therefore has negative Minkowski product with $(\sinh\vte,\cosh\vte)\in W_1$. So we conclude that for $0\leq\la\leq\pi$, there holds ${\rm Im}\,p(\vte+i\la)\cdot x\leq 0$.

Hence
\begin{align*}
 \left|f^-_{p_\perp}(\vte+i\la)\right| &\leq \int d^dx\, |f(x)|e^{{\rm Im}\,p(\vte+i\la)\cdot x} \leq \int d^dx\, |f(x)| = \|f\|_1\,,\qquad p_\perp\in\Rl^{d-2}\,.
\end{align*}
By using the three lines theorem as in \cite{GL07-1}, it can also be shown that $f^-_{p_\perp}$ is not only bounded on the strip $0\leq\la\leq\pi$, but moreover satisfies
\begin{align}
\int d\vte\,|f^-_{p_\perp}(\vte+i\la)|^2 \leq
\int d\vte\,|f^-_{p_\perp}(\vte)|^2,\qquad 0\leq\la\leq\pi\,. 
\end{align}
The same inequality holds for $g^+_{p_\perp}(\vte+i\la)$ because the support of $g$ lies in $W_1'=-W_1$. Since $\cosh(\vte+i\pi)=-\cosh\vte$ and $\sinh(\vte+i\pi)=-\sinh\vte$, the boundary values of these functions at $\la=\pi$ are given by, $\vte\in\Rl$, $p_\perp\in\Rl^{d-2}$,
\begin{align}
 f^-_{p_\perp}(\vte+i\pi) &= f^+_{-p_\perp}(\vte)\,, \qquad g^+_{p_\perp}(\vte+i\pi) = g^-_{-p_\perp}(\vte)\,.\label{g-bnd}
\end{align}
We now study the behaviour of the (entire analytic) function $\vte\mapsto e^{ip(\vte)Q_1q}$ appearing in \eqref{phicomm}. Using the general form of $Q_1$ \eqref{q1-d}, we find
\begin{align}
 {\rm Im}\,p(\vte+i\la)Q_1q = \kae\, m_\perp\sin\la\, 
\left(
\begin{array}{cc}
 \cosh\vte\\\sinh\vte
\end{array}
\right)
\cdot
\left(
\begin{array}{cc}
 q_0\\ q_1
\end{array}
\right)\geq 0\,.
\end{align}
This expression is positive since $\kae\, m_\perp\sin\la\geq 0$ for $0\leq\la\leq\pi$ and the vectors $(\cosh\vte,\sinh\vte)$ and $(q_0,q_1)$ both lie in the two-dimensional forward lightcone and therefore have a positive Minkowski product.

Hence the exponential factor $e^{ipQ_1q}$ is bounded on the strip $0\leq\la\leq\pi$, $\vte\in\Rl$, by $|e^{ip(\vte+i\la)Q_1q}|\leq 1$. Together with the analyticity and boundedness properties of the functions $f^-_{p_\perp}$ and $g^+_{p_\perp}$, this implies that we can shift the contour of the $\vte$-integration in \eqref{phicomm} from $\Rl$ to $\Rl+i\pi$. Since the boundary values at ${\rm Im}\,\vte=\pi$ are given by \eqref{g-bnd}, we arrive at
\begin{align*}
\int d\mu(p)\,e^{ipQ_1q} f^-(p) g^+(p)
&=
\int d^{d-2}p_\perp\int d\vte\,e^{ip(\vte)Q_1q} f^-_{p_\perp}(\vte)g^+_{p_\perp}(\vte)
\\
&=
\int d^{d-2}p_\perp\int d\vte\,e^{ip(\vte+i\pi)Q_1q} f^+_{-p_\perp}(\vte)g^-_{-p_\perp}(\vte)\\
&=
\int d\mu(p)\,e^{-ipQ_1q}f^+(p)g^-(p)\,,
\end{align*}
implying that the field commutator \eqref{phicomm} vanishes.
{\hfill  $\square$ \\[2mm]}

We now have established the main properties of the fields $\phi_W$ as a covariant system relatively wedge-local quantum fields. The relevance of this observation for quantum field theories on noncommutative spacetimes is discussed in our conclusions.

\section{Two-particle interactions and local observables}\label{sec:qft}

In this section we discuss the model defined by the fields $\phi_W$ in the setting of quantum field theories on the usual,  ``commutative'' Minkowski spacetime $\Rl^d$, i.e. we here take the appearance of the matrices $Q$ in the underlying algebra \eqref{aQp} just as part of the definition of the quantum fields, without connection to noncommutative Minkowski space. We consider the two-dimensional case first.
\\\\
If the spacetime dimension is $d=2$, the sets $\W_0$ and $\Q$ consist only of two elements each,
\begin{align}
 \W_0=\{W_1,-W_1\}\,,\qquad \Q=\{Q_1,-Q_1\}\,,
\end{align}
where 
\begin{align}
 Q_1=\kae\,\left(\begin{array}{rr}
            0&+1\\-1&0
           \end{array}
\right)\,.
\end{align}
Fixing $\kae>0$, the isomorphism $Q$ \eqref{mapQ} between the homogeneous spaces $\W_0$ and $\Q$ is given by $Q(\pm W_1)=\pm Q_1$ (Lemma \ref{lemma:QW}).

In two dimensions, it it convenient to take the rapidity $\vte$ as the variable on the (one-dimensional) mass shell, i.e. we parametrize $H_m^+$ by $p(\vte)=m(\cosh\vte,\sinh\vte)$. With this notation, the exchange factors in the relations \eqref{aQp} are
\begin{align}\label{def:S2}
 e^{ip(\vte_1)Q_1p(\vte_2)} = e^{im^2\kae\sinh(\vte_1-\vte_2)} =: S_2(\vte_1-\vte_2)\,.
\end{align}
It is interesting to note that the so defined function $S_2$ is a {\em scattering function} in the sense of \cite{GL07-1}, i.e. $S_2$ is analytic and bounded in the strip\footnote{The function $S_2$ defined in \eqref{def:S2} is in fact entire analytic. However, for generic scattering functions, only analyticity in the strip $S(0,\pi)$ is required.} $S(0,\pi):=\{\zeta\in\Cl\,:\,0<{\rm Im}\,\zeta<\pi\}$, and furthermore satisfies the equations
\begin{align}\label{propS2}
 S_2(\vte)^{-1} = \overline{S_2(\vte)} = S_2(-\vte) = S_2(\vte+i\pi)\,,\qquad \vte\in\Rl\,. 
\end{align}
These relations are familiar from the context of factorizing S-matrices, where they express the fundamental properties of unitarity, hermitian analyticity and crossing symmetry of the scattering operator $S$ associated to $S_2$ \cite{smirnov-book,iagolnitzer-buch}. Moreover, $S$ trivially satisfies the Yang-Baxter equation since we are dealing here with only a single species of particles, i.e. it has all properties of an S-matrix of a completely integrable relativistic quantum field theory \cite{smirnov-book}.
\\\\
The relation of the construction carried out in the previous sections to an integrable quantum field theory model with scattering function $S_2$ becomes apparent when introducing the notations
\begin{align}
 z(\vte) &:= a(-Q_1,p(\vte))\,,&\quad \zd(\vte)&:=a^*(-Q_1,p(\vte))\,,\\ 
 z(\vte)'&:= a(Q_1,p(\vte))\,,&\quad  \zd(\vte)'&:=a^*(Q_1,p(\vte))\,.
\end{align}
Using the algebra \eqref{aQp}, we find the commutation relations
\begin{subequations}\label{zf1}
\begin{align}
 z(\vte_1)z(\vte_2)     &= S_2(\vte_1-\vte_2) \,z(\vte_2)z(\vte_1)\,,\\
 \zd(\vte_1)\zd(\vte_2) &= S_2(\vte_1-\vte_2) \,\zd(\vte_2)\zd(\vte_1)\,,\\
 z(\vte_1)\zd(\vte_2)   &= S_2(\vte_2-\vte_1) \,\zd(\vte_2)z(\vte_1) + \delta(\vte_1-\vte_2)\cdot 1\,,
\end{align}
\end{subequations}
i.e. the $z^\#(\vte)$ form a representation of the Zamolodchikov-Faddeev algebra \cite{Zamolodchikov:1978xm,Faddeev:1984si} with scattering function $S_2$. This algebra was also obtained by one of the authors (H.~G.) in \cite{Grosse-ZZ}. In the context of the relations \eqref{algahat}, the appearance of Zamolodchikov's algebra has already been noted by Kulish \cite{Kulish:2006jq}.

Similarly, the distributions $z^\#(\vte)'$ give
\begin{subequations}\label{zf2}
 \begin{align}
 z(\vte_1)'z(\vte_2)'    &= S_2(\vte_1-\vte_2)^{-1} \,z(\vte_2)'z(\vte_1)'\,,\\
 \zd(\vte_1)'\zd(\vte_2)'&= S_2(\vte_1-\vte_2)^{-1} \,\zd(\vte_2)'\zd(\vte_1)'\,,\\
 z(\vte_1)'\zd(\vte_2)'  &= S_2(\vte_2-\vte_1)^{-1} \,\zd(\vte_2)'z(\vte_1)' + \delta(\vte_1-\vte_2)\cdot 1\,, 
 \end{align}
\end{subequations}
and therefore form a representation of the Zamolodchikov-Faddeev algebra with scattering function $S_2^{-1}$. Moreover, these two representations lie in a particular relative position to each other, described by (cf. \eqref{aQp})
\begin{subequations}\label{zf12}
\begin{align}
 z(\vte_1)z(\vte_2)'    &= z(\vte_2)'z(\vte_1)\,,\label{comp1}\\
 \zd(\vte_1)\zd(\vte_2)'&= \zd(\vte_2)'\zd(\vte_1)\,,\label{comp1b}\\
 z(\vte_1)\zd(\vte_2)'  &= \zd(\vte_2)'z(\vte_1) + \delta(\vte_1-\vte_2)\cdot e^{-ip(\vte_1)Q_1P}\,.\label{comp2}
\end{align}
\end{subequations}
The algebraic structure summarized in (\ref{zf1}, \ref{zf2}, \ref{zf12}) lies at the root of a recent construction of quantum field theories with factorizing S-matrices \cite{Schroer:1997cq,Lechner:2003ts,Buchholz:2004qy,GL07-1}, initiated by Schroer \cite{Schroer:1997cq}. We briefly outline it here.

In this construction, one starts from a scattering function $S_2$ characterized by the conditions \eqref{propS2} and the above mentioned analyticity and boundedness properties. In particular, \eqref{def:S2} is an admissible choice for $S_2$. Having fixed the scattering function, one considers the algebra (\ref{zf1}, \ref{zf2}, \ref{zf12}), with the operator $\exp(ip(\vte_1)Q_1P)$ in \eqref{comp2} generalized to the second quantized, $S_2$-dependent multiplication operator $M(\vte)$ \cite{Lechner:2003ts},
\begin{align}
 (M(\vte)\Psi)_n(\vte_1,...,\vte_n) = \prod_{k=1}^n S_2(\vte_k-\vte)\cdot \Psi_n(\vte_1,...,\vte_n)\,.
\end{align}
To the algebras of the $z^\#(\vte)$ and $z^\#(\vte)'$, respectively, one associates the quantum fields \cite{Schroer:1997cq}
\begin{align}
 \phi(x) := \int d\vte\,\left(\zd(\vte)e^{ip(\vte)\cdot x}+z(\vte)e^{-ip(\vte)\cdot x}\right)
\end{align}
and \cite{Lechner:2003ts}
\begin{align}
 \phi'(x) := \int d\vte\,\left(\zd(\vte)'e^{ip(\vte)\cdot x}+z(\vte)'e^{-ip(\vte)\cdot x}\right)\,.
\end{align}
By exploiting the ``compatibility conditions'' \eqref{zf12} between the representations $z$ and $z'$, it was shown in \cite{Lechner:2003ts} that $\phi$ and $\phi'$ are relatively wedge-local to each other, i.e. $\phi(f)$ commutes with $\phi(g)$ if $-W_1+\supp f$ is spacelike separated from $W_1+\supp g$. In the present context, this is a special case of Proposition \ref{prop:phiW}. (Note that the invariant measure $d\mu(p)$ becomes Lebesgue measure $d\vte$ in the rapidity variable.)

Using operator-algebraic arguments developed in \cite{Buchholz:2004qy}, it was shown in \cite{GL07-1} that if $S_2$ is a {\em regular scattering function} in the sense that $\zeta\mapsto S_2(\zeta)$ can be continued to a bounded analytic function on a strip of the form $-\beta<{\rm Im}\,\zeta<\pi+\beta$, with $\beta>0$, then there exist local observables in the corresponding model theory. More precisely, for any open region $\OO\subset\Rl^2$ one finds in this case an infinite-dimensional von Neumann algebra $\A(\OO)$ of bounded operators\footnote{More precisely, if $\OO$ is a double cone, $\A(\OO)$ can be shown to be isomorphic to the hyperfinite type III$_1$ factor.} $A$ localized in $\OO$, i.e.
\begin{align}
 [A,\phi(f)]  &=0\qquad {\rm for}\quad\OO\subset (-W_1+\supp f)'\,,\\
 [A,\phi'(f)] &=0\qquad {\rm for}\quad\OO\subset (W_1+\supp f)'\,,\\
 [A_1,A_2]    &=0\qquad {\rm for}\quad A_1\in\A(\OO_1),A_2\in\A(\OO_2)\,,\quad \OO_1\subset\OO_2'\,.
\end{align}
These operator algebras satisfy all the properties that algebras of local observables have in a well-behaved quantum field theory \cite{GL07-1}. Furthermore, they can be used in the framework of Haag-Ruelle scattering theory \cite{araki} to compute the S-matrix of the model theory, which is found to be the one given by the scattering function $S_2$ \cite{GL07-1}.
\\\\
Also in the context of the model studied here, one may interpret the algebras generated by the wedge-local fields $\phi_W$ and their relative commutants in the same way as indicated above. However, it has to be noted that all the mentioned results pertaining to the existence and properties of {\em local} observables have been proven only for scattering functions which are regular in the sense described before. The scattering function \eqref{def:S2} is {\em not} regular, since $|S_2(\zeta)|$ is superexponentially increasing for $-\pi<{\rm Im }\,\zeta<0$, ${\rm Re}\,\zeta\to\pm\infty$. Therefore the analysis of \cite{GL07-1} does not apply here, i.e. the existence theorems for local observables established there possibly do not hold in the case at hand. This possible absence of strictly local observables may perhaps even be expected from the relation of our construction to non-commutative Minkowski space. 
\\\\
Nonetheless, the structural similarity to two-dimensional models with regular scattering functions suggests that the exponential terms $\exp(ipQq)$ appearing in the exchange relations of the fields $\phi_W$ are connected to scattering amplitudes. As our next step, this conjecture will be verified in the higher-dimensional context.
\\\\
Following the model-independent analysis of collision processes in wedge-local theories carried out by Borchers, Buchholz and Schroer \cite{Borchers:2000mz}, we now investigate two-particle scattering in the model defined by the fields $\phi_W$.

The properties of the fields $\phi_W(f)$ which are crucial in this context are the following, $W\in\W_0$, $f\in\Ss(\Rl^d)$:
\begin{itemize}
 \item $\phi_W(f)$ and $\phi_W(f)^*$ are closed operators containing the vacuum vector $\Om$ in their domains, and $\phi_W(f)\Om$, $\phi_W(f)^*\Om$ are single particle states.
\item $\phi_W(f)$ is localized in the wedge $(W+\supp f)''$.
\item $\phi_W(f)$ is {\em temperate} in the sense that 
\begin{align}
 x\longmapsto \phi_W(f)U(x)\Psi\,,\qquad\Psi\in\DD\,,
\end{align}
is continuous and bounded, $\|\phi_W(f)U(x)\Psi\|\leq c_{f,\Psi}$.
\end{itemize}

These three properties constitute the definition of a so-called {\em temperate polarization-free generator} \cite{Schroer:1997cq,Borchers:2000mz}. It has been shown in \cite{Borchers:2000mz} how such operators can be used to calculate two-particle scattering amplitudes in Haag-Ruelle collision theory, and we briefly review this procedure here.

For $t\in\Rl$ and $f\in\Ss(\Rl^d)$, let $f_t$ be defined by
\begin{align}\label{def:ft}
 f_t(x)	&:= (2\pi)^{-d/2}\int d^dp\,\fti(p)e^{ip\cdot x}e^{i(\omega_\sbp-p_0)t}\,.
\end{align}
For asymptotic $t\to\pm\infty$, the support of $f_t$ is essentially contained in $t\,\VV(f)$, where
\begin{align}
 \VV(f) := \{(1,\bp/\omega_\sbp)\,:\,p\in\supp\fti\,\}
\end{align}
is the velocity support of $f$ \cite{hepp-lsz,Borchers:2000mz} 

For test functions $f,g\in\Ss(\Rl^d)$ whose Fourier transforms have compact supports concentrated about points on the upper mass shell, we introduce a family of order relations $\prec_W$, $W\in\W_0$, by
\begin{align}
 f\prec_W g \;:\Longleftrightarrow\;\VV(g)-\VV(f)\subset W\,.
\end{align}
In a local quantum field theory with temperate polarization-free generators, the wedge-localization of these objects on the one hand, and the essential support of the test functions \eqref{def:ft} on the other hand, can be used to show that $\phi_W(f_t)\phi_{W'}(g_t)\Om$ converges (strongly) to incoming respectively outgoing two-particle scattering states for $t\to\pm\infty$ \cite{Borchers:2000mz}. For this result to hold it is important that the velocity supports of $f$ and $g$ are ordered with respect to the wedge $W$ in such a way that the essential localization regions of $\phi_W(f_t)$ and $\phi_{W'}(g_t)$ are far apart and spacelike in the limit $t\to\pm\infty$. Using the standard notation for collision states, one obtains for testfunctions $f,g$ with disjoints momentum supports in small neighbourhoods of points on the upper mass shell
\begin{align}
\lim_{t\to\infty}\phi_W(f_t)\phi_{W'}(g_t)\Om &= (\phi_W(f)\Om \times \phi_{W'}(g)\Om)\oout & &{\rm if}\qquad g\prec_W f\,,\\
\lim_{t\to-\infty}\phi_W(f_t)\phi_{W'}(g_t)\Om &= (\phi_W(f)\Om \times \phi_{W'}(g)\Om)\iin & &{\rm if}\qquad f\prec_W g\,.
\end{align}
In the model-independent setting, these formulas were derived with the help of an underlying local theory. However, since they involve only wedge-local quantities, they apply in principle also to the present wedge-local model. But since we have no a priori information about the local structure of this model, and in fact expect non-local features, it might happen that the limits $\lim_t\phi_W(f_t)\phi_{W'}(g_t)\Om$ do not exhibit all properties which the scattering states of a quantum field theory fully complying with the principle of locality have. Below we will see that in our model, these limits depend not only on the single particle states $\phi_W(f)\Om=f^+$ and $\phi_{W'}(g)\Om=g^+$, but also on the wedge $W$ -- in contradistinction to the situation encountered in a local theory.

In the model at hand, $\phi_W(f_t)\phi_{W'}(g_t)\Om$ is actually independent of $t\in\Rl$, since $\phi_W(f_t)=a^*(Q(W),{f_t}^+)$ in view of the support properties of $\fti$, and ${f_t}^+=f^+$ in view of \eqref{def:ft} and the definition of $f^+$. So we arrive at the following explicit form of two-particle scattering states
\begin{align}
(f^+ \times g^+)^W\oout &= a^*(Q(W),f^+)a^*(Q(W'),g^+)\Om & &{\rm if}\qquad g\prec_W f\,,\\
(f^+ \times g^+)^W\iin &= a^*(Q(W),f^+)a^*(Q(W'),g^+)\Om & &{\rm if}\qquad f\prec_W g\,.
\end{align}
We introduced here an extra index $W$ on the scattering states to indicate the dependence of $(f^+ \times g^+)^W_{\rm in/out}$ on the wedge with respect to which the test functions are ordered. Explicitely we find for $g\prec_W f$
\begin{align}
 (f^+\times g^+)^W\oout  &= \int d\mu(p)\int d\mu(q)\, f^+(p)g^+(q) e^{-\frac{i}{2}pQ(W)q}\,a^*(p)a^*(q)\Om\,,\\
 (f^+\times g^+)^W\iin &= \int d\mu(p)\int d\mu(q)\, f^+(p)g^+(q) e^{+\frac{i}{2}pQ(W)q}\,a^*(p)a^*(q)\Om\,.
\end{align}
For $Q_1=0$, the scalar products of these two-particle states coincide with the simple scattering amplitudes of an interaction-free theory. But for $Q_1\neq 0$, our model exhibits non-trivial S-matrix elements.

In $d>2$ dimensions, given two testfunctions $f,g$ with momentum supports about points on the mass shell, in general there exists a large family of wedges $W$ satisfying $g\prec_W f$. Since the two-particle collision states calculated above depend in a nontrivial manner on the choice of this wedge, we conclude that the scattering of our model is not governed by local fields if $d>2$.
\\\\
For a proper analysis of the status of local fields/observables, it is highly advantageous to use operator-algebraic techniques \cite{haag}. This will be done in a forthcoming paper, where it is also shown that the present model is only a particular example of a large class of theories\footnote{H. Grosse and G. Lechner, work in progress}. Another special model of this class has already been investigated by Buchholz and Summers \cite{Buchholz:2005xj}. All the models of this class have certain non-local properties in common. In particular, it can be derived from \cite{Borchers:2000mz} that they all violate the Reeh-Schlieder property and do not have the rich content of local observables characterizing a typical local quantum field theory.

\section{Conclusions and open questions}\label{sec:end}

In this article, we investigated a new model of wedge-localized quantum fields, which can be interpreted in two different ways: 

First, it may be viewed as a theory containing an infinity of models of free fields $\phi(Q,x)$ on noncommutative Minkowski spaces with different noncommutativity parameters $Q$. A representation of the proper, untwisted Poincar\'e group was used to implement the relativistic symmetry. This symmetry was found to be respected in the complete model, whereas in the subtheories given by fixed $Q$, the Lorentz group is broken down to its subgroup ${\rm SO}(1,1)\times{\rm SO}(d-2)$.

Most surprising, we found that the non-local fields $\phi(Q,x)$ satisfy clearcut relative localization properties, namely, they are localized in wedges. This new observation shall be studied more thoroughly elsewhere, and possibly applied to the construction of new models on noncommutative spacetimes.
\\
\\
The second possible perception of our model is to view it as a non-local, but wedge-local, quantum field theory on ordinary Minkowski space. From this point of view, we analyzed collision processes, which were found to lead to non-trivial S-matrix elements.

However, the emerging interacting quantum field theory is not generated by local observables. In fact, the existence of any non-trivial strictly local observables in this model is presently an open question. This non-local structure can be understood from the well-known incompatibility of local interacting quantum fields in higher dimensions, and a simple form of the interaction, as in integrable models.

For non-local theories, this incompatibility does not exist \cite{Baumgartel:1984ma}. It might therefore be worthwhile to investigate non-local quantum field theories which still satisfy certain weak localization properties, and interact in a manner similar to completely integrable models. In fact, such models might exist as the effective description of field theories on noncommutative spacetimes.

\acknowledgments
We would like to thank S.~Armstrong for helpful discussions about group actions and D.~Buchholz for the communication of references regarding non-local quantum field theories. The hospitality of the Erwin Schr\"odinger Institute is gratefully acknowledged.

\bibliography{/home/gandalf/Physik/Artikelsammlung/Bibtex-Database.bib}
\bibliographystyle{JHEP}

\end{document}